%Paper: hep-th/9406042
%From: thk@fwi.uva.nl (Tom H. Koornwinder )
%Date: Wed, 8 Jun 1994 17:22:49 +0200

\magnification=\magstephalf
\baselineskip=13pt
\parindent=20pt
\input amssym.def
\newsymbol\boxtimes 1202
%---------------------------------
%\input rapmcr
%----------------------------------
\def\ifundefined#1{\expandafter\ifx\csname#1\endcsname\relax}

\newif\ifdevelop \developfalse

\newtoks\chnumber
\newtoks\sectionnumber
\newcount\equationnumber
\newcount\thnumber

\def\assignnumber#1#2{%
	\ifundefined{#1}\relax\else\message{#1 already defined}\fi
	\expandafter\xdef\csname#1\endcsname
 {\if-\the\sectionnumber\else\the\sectionnumber.\fi\the#2}%
	}%

\hyphenation{pre-print wave-lets}

\def\br={\hfil\break}

\def\bLP{\bigbreak\noindent}

\font\chstil=cmbx10 at 12truept
\def\beginchapter #1 #2 {\vskip0pt plus.1\vsize
 \penalty-250\vskip0pt plus-.1\vsize
          \bigbreak\bigskip
          \chnumber{#1} \noindent{\chstil Chapter #1.\ \  #2}\par
           \nobreak\medskip\noindent}
\def\beginsektion #1 #2 {\ifdevelop\vfill\eject\pageno1\else
	\vskip0pt plus.1\vsize\penalty-250\vskip0pt plus-.1\vsize
	\bigbreak\bigskip\fi
	\sectionnumber{#1} \equationnumber0\thnumber0
 \noindent{\bf\if-#1\else #1. \fi#2}\par
	\nobreak\medskip\noindent}
\def\Subsec#1 #2 {\bLP{\bf
 \if-\the\sectionnumber\else\the\sectionnumber.\fi#1. #2.}\quad}
\def\Appendix#1 #2 {\ifdevelop\vfill\eject\pageno1\else
	\vskip0pt plus.1\vsize\penalty-250\vskip0pt plus-.1\vsize
	\bigbreak\bigskip\fi
	\sectionnumber{#1} \equationnumber0\thnumber0
	\leftline{\bf Appendix #1. #2}%
	\nobreak\smallskip\noindent}

\def\today{\ifcase\month\or January\or February\or March\or April\or May\or
 June\or July\or August\or September\or October\or November\or December\fi
 \space\number\day, \number\year}

\def\nologo{\expandafter\let\csname logo\string @\endcsname=\empty}
\def\:{:\allowbreak }

\def\ep{\varepsilon}

\def\De{\Delta}

\def\CC{{\Bbb C}}

\def\RR{{\Bbb R}}

\def\FSA{{\cal A}}

\def\FSH{{\cal H}}

\def\id{{\rm id}}

\def\tr{{\rm tr}}

\def\Span{{\rm Span}}

\let\ten=\otimes

\def\eq#1{\relax
	\global\advance\equationnumber by 1
	\assignnumber{EN#1}\equationnumber
	\ifdevelop{\rm(#1)}%
	\else{\rm (\csname EN#1\endcsname)}\fi%
	}

\def\eqtag#1{\relax\ifdevelop(#1)\else\eqtagelse{#1}\fi}

\def\eqtagelse#1{\ifundefined{EN#1}\message{EN#1 undefined}{\sl (#1)}%
	\else(\csname EN#1\endcsname)\fi%
	}

\def\thname#1{\relax
	\global\advance\thnumber by 1
	\assignnumber{TH#1}\thnumber
	\ifdevelop#1%
	\else\csname TH#1\endcsname\fi%
	}

\def\beginth#1 #2 {\bigbreak\noindent{\bf #1\enspace \thname{#2}}
         ---\hskip4pt}
\def\thtag#1{\relax\ifdevelop#1\else\thtagelse{#1}\fi}

\def\thtagelse#1{\ifundefined{TH#1}\message{TH#1 undefined}{\sl #1}%
	\else[\csname TH#1\endcsname]\fi}

%----------------------------------------------------
%\input macr
%---------------------------------------------------
\font\tx=cmr10
\font\thstil=cmsl10
\font\titlestil=cmr17
\font\authorstil=cmr12
\font\small=cmr9
\font\typew=cmtt9
\font\smallsl=cmsl9
\font\defstil=cmsl10

\def\RR{{\Bbb R}}
\def\CC{{\Bbb C}}

\def\id{\hbox{\tx id}}
\def\dd{\hbox{\tx d}}

\def\Span{\hbox{\tx Span}}

\def\tr{\hbox{\tx tr}}
\def\a{\alpha}

\def\tr{\hbox{\tx tr}}
\def\innp#1 #2 {\langle #1 , #2 \rangle}

\def\ten{\otimes}

\def\eps{\varepsilon}

\def\ltextindent#1{\hbox to \hangindent{#1\hss}\ignorespaces}

\def\nextsektion{\goodbreak}

\newskip\proofskipamount
\proofskipamount=8pt plus 2pt minus 2 pt
\def\proofskip{\vskip\proofskipamount}
\def\endth{\par \ifdim\lastskip<\bigskipamount \removelastskip\penalty55
\bigskip\fi}
\def\beginproof{\removelastskip\penalty55\proofskip\noindent}
\def\endproof{\bigbreak}

\def\newtheme{\bigbreak}

\def\titleone#1\par{{\titlestil \centerline{#1}\bigbreak}}
\def\titletwo#1\par{\removelastskip\medskip
                     \centerline{{\titlestil #1}}\bigbreak}
\def\authors#1\par{\bigbreak{\authorstil\centerline{#1}}\bigbreak}
\font\curs=cmsl10
\font\kleinfett=cmbx10 at 10 truept
\def\CWI{\bigbreak\centerline{\curs ${}^1$CWI}
     \centerline{\curs P.O.\ Box 94079}
      \centerline{\curs 1090 GB  Amsterdam, The Netherlands}
     \bigbreak}
\def\UvA{\medbreak\centerline{\curs ${}^2$University of Amsterdam}
     \centerline{\curs Faculty of Mathematics and Computer Science}
     \centerline{\curs Plantage Muidergracht 24}
     \centerline{\curs 1018 TV Amsterdam, The Netherlands}
     \medbreak}
\def\email#1\par{\smallbreak{\small \centerline{email: {\typew #1}}
    \bigbreak}}
\def\abstract#1\par{\bigbreak \centerline{\kleinfett Abstract}\smallbreak
      {\baselineskip=12pt\leftskip=1.5\parindent
     \rightskip=1.5\parindent
     \noindent{\small #1}\medbreak}}
\def\key#1\par{{\baselineskip=12pt\leftskip=1.5\parindent
     \rightskip=1.5\parindent
     \noindent{\smallsl Keywords and Phrases:\ }{\small #1}\par}\vskip 0.7cm}
\def\class#1\par{\medbreak{\baselineskip=12pt\leftskip=1.5\parindent
     \rightskip=1.5\parindent
     \noindent{\smallsl AMS Subject Classification (1991):}
     {\small #1}\par}}

\font\eightrm=cmr9
\long\def\fussnote#1#2{{\baselineskip=10pt
      \setbox\strutbox=\hbox{\vrule height 7pt depth 2pt width 0pt}%
      \eightrm
      \everypar{\hangindent=\parindent}%
      \footnote{$^{#1}$}{#2}\everypar{}}}

%---------------------------------------------------------
\def\phi{\varphi}
%--------------------------------------------------------
%\input ndrefdef
%----------------------------------------------------------
\def\Arv{[1]}
\def\Bonneau{[2]}
\def\BourbakiAlgeight{[3]}
\def\msdth{[4]}
\def\Effros{[5]}
\def\Gerst{[6]}
\def\Guichardet{[7]}
\def\Koorlect{[8]}
\def\Kustermans{[9]}
\def\Maclane{[10]}
\def\FRT{[11]}
\def\Sweedler{[12]}
\def\Vaksoib{[13]}
\def\Wang{[14]}
\def\Wangdis{[15]}
\def\Worcomp{[16]}
\def\Wortan{[17]}
\def\Worrem{[18]}
\def\Worprep{[19]}
%---------------------------------------------------------------
%\input ndsec0
%---------------------------------------------------------------
\titleone  CQG Algebras: a Direct Algebraic Approach

\titletwo  to Compact Quantum Groups

\authors Mathijs S.\ Dijkhuizen${}^1$ and Tom H.\ Koornwinder${}^2$

\CWI\UvA
\email thijs@cwi.nl \quad thk@fwi.uva.nl

\abstract
The purely algebraic notion of CQG algebra (algebra of functions
on a compact quantum group) is defined. In a straightforward
algebraic manner, the Peter-Weyl theorem for CQG algebras and the
existence of a unique positive definite Haar functional on any
CQG algebra are established. It is shown that a CQG algebra can be
naturally completed to a $C^\ast$-algebra. The relations between our
approach and several other approaches to compact quantum groups
are discussed.

\class 81R50, 16W30, 22C05.

\key compact quantum group, Hopf $\ast$-algebra, unitary corepresentation,
 CQG algebra, Haar functional, Peter-Weyl theorem, $C^\ast$-algebra completion.

\def\Supnormtheorem{4.4}
\beginsektion 0 {Introduction}
Compact quantum groups are fairly well understood, on the one hand for
special cases such as $SU_q(2)$ (cf.\ the early paper \Vaksoib) and, more
generally, quantum analogues of the classical compact Lie
groups  and beyond (cf.\ \FRT), and on the other hand in a general theory
started by
Woronowicz \Worcomp, \Wortan, \Worrem\
for compact matrix quantum groups.
A crucial aspect of Woronowicz's general theory is the existence theorem
for a positive Haar functional. Some $C^\ast$-algebra theory is
used in the demonstration of that theorem, and, actually, a $C^\ast$-algebra
is already present in Woronowicz's definition of compact matrix quantum group.
This is in contrast with the special cases, where a Hopf $\ast$-algebra is
presented as an algebra by generators and relations. The demonstration of
a $C^\ast$-completion for such explicit algebras can be quite cumbersome,
and is actually not necessary for many applications where one is only
interested in algebraic aspects. Thus, in the special cases one usually
develops the theory in an ad hoc algebraic manner, and in this way one arrives
at
results fitting into Woronowicz's general theory without actually invoking
his theorems.
By the way, also in the general case the main results of Woronowicz's
theory (Schur orthogonality relations, Peter-Weyl theorem) can be formulated
in a meaningful way on the algebraic level.

In this note we propose a purely algebraic approach to general
compact quantum groups. After some preliminaries in \S1,
we define, in \S2, a CQG algebra (associated with a
compact quantum group) as a Hopf $\ast$-algebra which is the linear span of
the matrix elements of its finite-dimensional unitary corepresentations.
A CMQG algebra (associated with a compact matrix quantum group)
is then a finitely generated CQG algebra.
These definitions and the subsequent
development of the theory (in \S3) do not involve $C^\ast$-algebras.
All main results of \Worcomp, as far as they are on the Hopf $*$-algebra level,
are thus proved in an algebraic way.
In particular, the existence of a unique (not a priori positive)
Haar functional on a CQG algebra is immediate, and its positivity and
faithfulness (on the CQG algebra) is one of the results of \S3.
We show in \S4 that a CQG algebra has a natural
$C^\ast$-completion, by which we make contact with \Worcomp.
The paper concludes (in \S5) with a comparison of various approaches to
compact quantum groups which have appeared in the literature.
Particular mention here deserves the paper \Effros\
by Effros \& Ruan, who earlier introduced
the same algebras as our CQG algebras, but called them differently and
also developed the theory in a different direction.
We also mention the paper \Bonneau (see also \Gerst), in which, among other
things, a notion
of so-called preferred deformation of the algebra of representative functions
on a compact connected Lie group is defined and a number of its properties
are studied. Although the techniques used in \Bonneau\ are quite different,
some of the results are similar in spirit to ours.

The results presented here are also part of the Ph.D.\ thesis of
Dijkhuizen \msdth, while a more tutorial presentation will appear in lecture
notes by Koornwinder \Koorlect.

\smallbreak\noindent
{\sl Acknowledgement}\quad
We thank P. Podle\'s for a helpful remark in connection with the proof of
Theorem \Supnormtheorem. Part of the research for this paper by the
second author was done at CWI, Amsterdam.
\nextsektion
%------------------------------------------------------------
%\input ndsec1
%----------------------------------------------------------
\beginsektion 1 {Preliminaries}
All vector spaces are taken over the field of complex numbers $\CC$.
All tensor products of vector spaces are algebraic unless explicitly
mentioned otherwise. We canonically identify the tensor products
$V\ten \CC$ and $\CC\ten V$ with $V$, for any vector space $V$.
For more information about the material treated in this section see
\Sweedler, \msdth, \Koorlect.

In a Hopf algebra $A$ we write
$\Delta\colon A\to A\ten A$ for the comultiplication,
$\eps\colon A\to \CC$ for the counit, and $S\colon A\to A$ for the antipode.
We recall the symbolic notation for $\Delta$ and its iterates:
$$\Delta(a) = \sum_{(a)} a_{(1)}\otimes a_{(2)}, \quad
(\Delta\otimes\id) \circ \Delta (a) = (\id\otimes \Delta)\circ \Delta (a) =
\sum_{(a)} a_{(1)} \otimes a_{(2)} \otimes a_{(3)}. \eqno\eq{symbol}$$
A {\defstil Hopf $\ast$-algebra}
is a Hopf algebra $A$ endowed with a conjugate linear
involutive
mapping $\ast\colon A\to A$ such that
$A$ as an algebra is a $\ast$-algebra and such that $\Delta$ and $\eps$ are
$\ast$-homomorphisms. It then follows  that the antipode $S$
is bijective and satisfies
$S\circ \ast \circ S \circ \ast = \id$.

Let $A$ be a Hopf algebra. A {\defstil corepresentation}
of $A$ in a vector space $V$
is a
linear mapping $\pi\colon  V\to V\otimes A$ such that
$$(\pi\otimes\id)\circ\pi = (\id\otimes\Delta)\circ\pi, \quad
(\id\otimes\varepsilon)\circ\pi = \id. \eqno\eq{comodule}
 $$
We shall sometimes use the following symbolic notation
for $\pi$ and its iterates:
$$\pi(v) = \sum_{(v)} v_{(1)} \otimes v_{(2)}, \quad
(\pi\otimes\id)\circ\pi(v) = (\id\otimes\Delta)\circ\pi(v)=
\sum_{(v)} v_{(1)} \otimes v_{(2)} \otimes v_{(3)}.
\eqno\eq{symbolmod}$$
Here $v\in V$, the $v_{(1)}$ are in $V$, and the $v_{(2)}, v_{(3)}$ in A.
If the corepresentation space $V$ is finite-dimensional and $\{v_i\}$
is a basis of
$V$, then we write $\pi(v_j) = \sum_i  v_i\otimes \pi_{ij}$, where the
$\pi_{ij}$ are
elements of $A$. Then $\pi=(\pi_{ij})$ is a matrix corepresentation of $A$:
$$\Delta(\pi_{ij}) = \sum_k \pi_{ik}\otimes \pi_{kj},
\quad \varepsilon(\pi_{ij}) = \delta_{ij}.
\eqno\eq{axiommatcorep}$$
Given two corepresentations $\pi$ in $V$ and $\rho$ in $W$, a linear
mapping $\phi\colon
V\to W$ is called an {\defstil intertwining operator} if
$\rho\circ \phi = (\phi\otimes\id)\circ \pi$.

Given a corepresentation $\pi$ of $A$ in a finite-dimensional vector space $V$,
the {\defstil contragredient
corepresentation} $\pi'$ of $\pi$ is the corepresentation of $A$
in the linear dual $V'$
defined by
$$(v\ten\id)\circ \pi'(v') = (v'\otimes S)\circ\pi(v),\quad
v'\in V',\; v\in V. \eqno\eq{contragcorep}$$
If we write $\pi=(\pi_{ij})$ with respect to a basis
$\{v_i\}$ of $V$ and $\pi'=(\pi'_{ij})$  with respect to the dual basis
$\{v^i\}$ of $V'$ then
$$\pi'_{ij} = S(\pi_{ji}). \eqno\eq{contragcocoeff}$$

Given two corepresentations $\pi$, $\rho$ in finite-dimensional vector
spaces $V$, $W$, respectively, their {\defstil
tensor product}
$\pi\boxtimes \rho$ is the corepresentation of $A$
in the vector space $V\ten W$ defined in symbolic notation by
$$\pi\boxtimes \rho (v\otimes w) = \sum_{(v), (w)} v_{(1)} \otimes w_{(1)}
\otimes
v_{(2)} w_{(2)}.\eqno\eq{tenscomodsymb}$$
If we write $\pi=(\pi_{ij})$ and $\rho=(\rho_{kl})$ with respect to a basis
$\{v_i\}$ of $V$ and $\{w_k\}$ of $W$ and if we write
$\pi\boxtimes \rho=((\pi\boxtimes \rho)_{ik,jl})$ with respect to the basis
$\{v_i\ten w_k\}$ of $V\ten W$ then
$(\pi\boxtimes \rho)_{ik,jl}=\pi_{ij}\,\rho_{kl}$.

Suppose $A$ is a Hopf $\ast$-algebra.
Let $V$ be a vector space endowed with an inner product. A corepresentation
$\pi$ of $A$ in
$V$ is called {\defstil unitary} if
$$\sum_{(v)} \innp {v_{(1)}} w \, S(v_{(2)}) =
\sum_{(w)} \innp v {w_{(1)}} \, w_{(2)}^\ast\quad  \forall v,w\in V.
\eqno\eq{unitardef}$$
Suppose  $V$ is finite-dimensional and $\pi=(\pi_{ij})$
with respect to an orthonormal basis $\{v_i\}$ of $V$.
Then $\pi$ is unitary if and only if the following equivalent conditions
are satisfied:
$$S(\pi_{ij}) = \pi_{ji}^\ast \iff \sum_k \pi_{ki}^\ast \pi_{kj} =
\delta_{ij}\, 1 \iff
\sum_k \pi_{ik} \pi_{jk}^\ast = \delta_{ij}\, 1. \eqno\eq{unitarmatdef}$$
A corepresentation $\pi$ in a vector space $V$ is called {\defstil
unitarizable}
if there exists an inner product on $V$ such that $\pi$ is unitary with
respect to this inner product.

The usual notions from representation theory such as direct sums,
invariant subspaces,
irreducibility, complete reducibility etc.\ all have an obvious
meaning in corepresentation
theory.
Note that a  unitary corepresentation always is completely reducible,
since the orthogonal
complement of an invariant subspace is again invariant.
\nextsektion
%--------------------------------------------------------------------
%\input ndsec2
%------------------------------------------------------------------
\beginsektion 2 {CQG algebras}
\beginth {Theorem} coeffhopfstar
{\thstil Let $A$ be a Hopf algebra.
Let $\Sigma$ denote the set of
equivalence classes of finite-dimensional irreducible
corepresentations of $A$.
For each $\a\in\Sigma$, let the matrix corepresentation
$\pi^\a=(\pi^\a_{ij})$ be a representative of the class $\a$ and
let $A_\a\subset A$ denote the span of its matrix
coefficients.
Then
$\sum_{\a\in\Sigma} A_\a$ is a direct sum and the $\pi_{ij}^\a$
are linearly independent.
\par}
\endth
\beginproof
This theorem is rather standard.
In 2.1.3, 1.1.33, 1.1.54 and 1.1.16 of \msdth\ it is pointed out that the
theorem is already valid in the case of finite-dimensional irreducible
corepresentations of a coalgebra, and that, by duality,
the proof can be reduced to a similar result for algebra representations,
see for instance \BourbakiAlgeight\ \S13, no.\ 3, pp.\ 154--155.
A self-contained proof not referring to algebra representations
but assuming that $S$ is invertible, is given in
\Koorlect\ Proposition 1.28.
\endproof
{}From now on we will work with a Hopf $\ast$-algebra $A$, and $\Sigma$ will
denote the set of
equivalence classes of finite-dimensional irreducible {\sl unitary}
corepresentations of $A$. For $\a\in\Sigma$, let $\pi^\a$ and $A_\a$ be as
in the above theorem, but the matrix corepresentation $(\pi^\a_{ij})$ is now
supposed to be unitary.
\beginth {Definition} cqgdef
A {\defstil CQG algebra} is a Hopf $\ast$-algebra which is spanned by the
coefficients of its finite-dimensional unitary (irreducible) corepresentations.
We then say that $A$ is the Hopf $\ast$-algebra  associated with
a compact quantum group.
The decomposition $A= \sum_{\alpha\in\Sigma} A_\a$ is called the
{\defstil Peter-Weyl decomposition} of $A$.
\endth
\beginth {Proposition} CQGirredunit
{\thstil Let $A$ be a CQG algebra. Then:
\smallskip
{\parindent=1cm
\item{(i)} Every finite-dimensional irreducible corepresentation of $A$
is equivalent
to some $\pi^\alpha$ ($\alpha\in \Sigma$).
\item{(ii)} Every finite-dimensional irreducible corepresentation of $A$
is unitarizable.
\item{(iii)} If $\pi$ is a finite-dimensional
unitarizable matrix corepresentation of $A$ then
so is its contragredient $\pi'$.
\par}
 \par}
\endth
\beginproof
Assertion (i) is an immediate consequence of \thtag{coeffhopfstar} and
\thtag{cqgdef}.
Assertion (ii) follows from (i). For (iii) observe that a finite-dimensional
unitarizable corepresentation $\pi$ is completely reducible,
hence so is $\pi'$.
\endproof
\beginth {Proposition} cmqgequiv
{\thstil For a Hopf $\ast$-algebra $A$ the following conditions are equivalent:
\smallskip
{\parindent=1cm
\item{(i)} $A$ is a finitely generated CQG algebra.
\item{(ii)} There is a finite-dimensional unitary corepresentation of $A$ whose
matrix coefficients generate $A$ as an algebra.
\item{(iii)} There is a finite-dimensional  corepresentation $\pi$ of $A$
such that both $\pi$ and $\pi'$ are unitarizable and such that $A$ is generated
as an algebra by the matrix coefficients of $\pi$ and $\pi'$.
\item{(iv)} There is a finite subset $\{\alpha_1, \ldots, \alpha_n\}
\subset \Sigma$
such that the matrix coefficients of the $\pi^{\alpha_i}$ ($1\leq i\leq n$)
generate
$A$ as an algebra.
\par}
\par}
\endth
\beginproof
Assume (i) and pick a finite set of generators of $A$.
By definition of a CQG algebra,
each of the generators
is a linear combination of matrix coefficients of finite-dimensional unitary
corepresentations of $A$. Taking the direct sum of all the corepresentations
involved, we obtain a finite-dimensional unitary corepresentation whose
matrix coefficients generate $A$. This proves (i)$\Rightarrow$(ii).
The implication
(ii)$\Rightarrow$(iii) follows from \thtag{CQGirredunit} (iii).
The implications (iii)$\Rightarrow$(iv) and (iv)$\Rightarrow$(i) are immediate.
This concludes the proof.
\endproof
\beginth {Definition} cmqgdef
 A CQG algebra is called a {\defstil CMQG algebra} if it satisfies the
equivalent conditions of \thtag{cmqgequiv}.
A CMQG algebra is said to be associated with a compact matrix quantum group.
\endth
It can be easily shown (cf.\ \msdth\ p.\ 57--60) that the category
of CQG algebras is closed under taking inductive limits (for a definition of
inductive limit see also \Maclane\ p.\ 67). The standard fact that any compact
group can be written as a projective limit of compact Lie groups generalizes
to the statement that any CQG algebra is the inductive limit of CMQG algebras.
Conversely, given a family of CQG algebras $(A_\lambda)_{\lambda\in\Lambda}$,
the
tensor products of finite subfamilies of $(A_\lambda)_{\lambda\in\Lambda}$
naturally form
an inductive family whose limit generally is a non-finitely generated CQG
algebra.
In this way, one can construct examples of non-finitely generated CQG algebras
starting from an infinite family of CMQG algebras.
\nextsektion
%------------------------------------------
%\input ndsec3
%------------------------------------------
\beginsektion 3 {The Haar functional}
We now discuss the concept of Haar functional and its relation to CQG algebras.
\beginth Definition Haar
Let $A$ be a Hopf $\ast$-algebra. A {\defstil Haar
functional} on $A$ is a linear functional $h\colon A\to \CC$
which satisfies
$h(1) = 1$,  and  is such that
$$(h\ten\id) \circ \Delta(a) = h(a)\, 1 = (\id\ten h) \circ \Delta(a),
\quad a\in A. \eqno\eq{haarinvar}$$
The invariance \eqtag{haarinvar} of a  Haar functional $h\colon A\to \CC$
with respect to comultiplication
can also be written as
$$\sum_{(a)} h(a_{(1)}) a_{(2)} = h(a)1 = \sum_{(a)} h(a_{(2)}) a_{(1)},
\quad a\in A.
\eqno\eq{explHaar}$$
A Haar functional $h\colon A\to \CC$ is called {\defstil positive} if
$h(a^* a)\ge0$ for all $a\in A$, and it is called {\defstil positive definite}
or {\defstil faithful} if $h(a^\ast a) > 0$ for all nonzero $a\in A$.
It can be easily proved that, if a Haar functional $h$ satisfies
$h(a^* a)\in\RR$ for all $a\in A$, then
$h(a^\ast) = \overline{h(a)}$
for all $a\in A$.
\endth
Let now $A$ be a CQG algebra.
Denote by 1 the unique $\alpha\in\Sigma$ such that $\pi^\alpha$ is
equivalent to the one-dimensional unitary matrix corepresentation $(1)$.
Also, for each $\alpha\in \Sigma$, let $\alpha'$ be the unique
$\beta\in \Sigma$ such that
$\pi^\beta$ is  equivalent to $(\pi^\alpha)'$, cf.\
\thtag{CQGirredunit} (iii).

We now define  a linear form
$h\colon A \to \CC$ by setting
$$h (a) = \cases{0 & if $a\in A_\alpha, \; \alpha \neq 1$, \cr
                 1 & if $a=1$. \cr}\eqno\eq{haardef}$$

\beginth Proposition  Haarprop
{\thstil Let $A$ be a CQG algebra and let $h\colon A\to \CC$
be the linear form defined in \eqtag{haardef}.
Then $h$ is a Haar functional on $A$ and satisfies
$h(S(a)) = h(a)$ and $h(a^\ast) = \overline{h(a)}$ for all $a\in A$.
Any linear functional $h'\colon A\to \CC$ such that $h'(1) = 1$ and
$(h'\otimes \id)\circ \Delta(a) = h'(a)1$ for all $a\in A$ is equal to $h$.
\par}
\endth
\beginproof
The proof is completely elementary.
\endproof
Let $G$ be a compact group and let $A$ be the corresponding CQG algebra
of representative functions on $G$. If $\dd x$ denotes the Haar measure on
$G$, then the Haar functional on $A$ is given by
$$h(a) = \int_G a(x) \dd x$$
and \eqtag{haarinvar} expresses the invariance of the Haar measure with respect
to group multiplication. It is well-known that the Haar measure is a positive
measure
and that the support of $\dd x$ is
equal to $G$, in other words, the Haar functional $h$ is positive definite.
We are going to prove the same result for a general CQG algebra. We first prove
an important lemma:
\beginth Lemma haarintertwin
{\thstil Let $A$ be a Hopf algebra  and suppose
$h\colon A\to \CC$ is a linear form satisfying \eqtag{haarinvar} and
such that $h(1)=1$. Let $\rho$ and $\sigma$ be matrix
corepresentations of $A$. Then
$$
\eqalignno{
\sum_l h(\sigma_{ij}S(\rho_{kl}))\rho_{lm}&=
\sum_l \sigma_{il}\,h(\sigma_{lj}S(\rho_{km})),&\eq{rhosigmaone}
\cr
\sum_l h(S(\rho_{ij})\sigma_{kl})\sigma_{lm}&=
\sum_l \rho_{il}\,h(S(\rho_{lj})\sigma_{km}).&\eq{rhosigmatwo}
\cr}
$$
With the notation
$$
A_{il}^{(j,k)}:=h(\sigma_{ij}S(\rho_{kl})),\quad
B_{il}^{(j,k)}:=h(S(\rho_{ij})\sigma_{kl}),\eqno\eq{ABrhosigma}
$$
the identities \eqtag{rhosigmaone} and \eqtag{rhosigmatwo} can be rewritten as
$$
A^{(j,k)}\rho=\sigma A^{(j,k)},\quad
B^{(j,k)}\sigma=\rho B^{(j,k)}.\eqno\eq{ABintertwin}
$$
Thus $A^{(j,k)}$ is an intertwining operator for $\rho$ and $\sigma$, and
$B^{(j,k)}$ is an intertwining operator for $\sigma$ and $\rho$.
\par}
\endth
\beginproof
For the proof of \eqtag{rhosigmaone} we write
$$
\eqalignno{
h(\sigma_{ij}S(\rho_{kl}))\,1=
(\id\ten h)(\Delta(\sigma_{ij}S(\rho_{kl})))=
\sum_{p,n}(\id\ten h)((\sigma_{ip}\ten \sigma_{pj})(S(\rho_{nl})\ten
S(\rho_{kn}))&
\cr
=\sum_{p,n} h(\sigma_{pj}S(\rho_{kn}))\sigma_{ip}S(\rho_{nl}).&
\cr}
$$
Substituting this equality in the left-hand side  of equation
\eqtag{rhosigmaone} and then using that
$\sum_l S(\rho_{nl})\rho_{lm}=\delta_{nm}\,1$, we arrive at
\eqtag{rhosigmaone}.
Similarly, \eqtag{rhosigmatwo} is obtained from a substitution of
$$
h(S(\rho_{lj})\sigma_{km})\,1=(h\ten\id)(\Delta(S(\rho_{lj})\sigma_{km}))
$$
in the right-hand side  of \eqtag{rhosigmatwo}.
\endproof
\beginth Proposition schurrelnoneq
{\thstil Let $A$ be a Hopf algebra and suppose
$h\colon A\to \CC$ is a linear form satisfying \eqtag{haarinvar} and
such that $h(1)=1$. If $\rho$ and $\sigma$ are non-equivalent
irreducible matrix corepresentations of $A$ then
$$h(\sigma_{kl} S(\rho_{ij})) =0, \quad h(S(\rho_{kl}) \sigma_{ij}) = 0.$$
\par}
\endth
\beginproof
We use the notation of \thtag{haarintertwin}. By the Schur lemma for
corepresentations,
\eqtag{ABintertwin} yields that $A^{(j,k)}_{il} = 0$ and $B^{(j,k)}_{il} = 0$.
Hence, by \eqtag{ABrhosigma}, the assertion.
\endproof
If $(\pi,V)$ is a finite-dimensional corepresentation of a Hopf algebra $A$,
then
the corepresentation space $V''$ of the double contragredient corepresentation
$\pi''$ of $\pi$ can be naturally identified with $V$.
\beginth Proposition schurreleq
{\thstil Let $A$ be a Hopf algebra with invertible antipode $S$ and suppose
$h\colon A\to \CC$ is a linear form satisfying \eqtag{haarinvar} and
such that $h(1)=1$ and $h(S(a))=h(a)$. Let $\rho$ be an irreducible matrix
corepresentation of $A$. Then $\rho$ is equivalent to its double contragredient
$\rho''$. Let $F$ be any invertible operator intertwining $\rho$ and
$\rho''$. Then $\tr(F) \neq 0$ and $\tr(F^{-1})\neq 0$ and
$$\eqalignno{& h(\rho_{kl} S(\rho_{ij})) = \delta_{kj} {F_{il} \over \tr(F)},
& \eq{schureqone}\cr
& h(S(\rho_{kl}) \rho_{ij}) = \delta_{kj} {(F^{-1})_{il}\over \tr(F^{-1})}.
& \eq{schureqtwo}\cr}$$
\par}
\endth
\beginproof
Putting $\sigma= \rho''$ in \thtag{haarintertwin} and using the fact that
$h(S(a)) = h(a)$, we deduce from \eqtag{rhosigmaone} and \eqtag{rhosigmatwo}
that
$$\sum_l h(\rho_{kl} S(\rho_{ij})) \rho_{lm}  = \sum_l S^2(\rho_{il})\,
h(\rho_{km} S(\rho_{lj})),$$
$$\sum_l h(S(\rho_{kl}) \rho_{ij}) S^2(\rho_{lm})  = \sum_l \rho_{il}\,
h(S(\rho_{km}) \rho_{lj}).$$
With the notation
$$\tilde A^{(j,k)}_{il} = h(\rho_{kl} S(\rho_{ij})), \quad
\tilde B^{(j,k)}_{il} = h(S(\rho_{kl}) \rho_{ij})\eqno\eq{tildeABrhosigma}$$
we have $\tilde A^{(j,k)} \rho = \rho'' A^{(j,k)}$ and
$\tilde B^{(j,k)} \rho'' =
\rho \tilde B^{(j,k)}$.
On the other hand, if we apply \thtag{haarintertwin}
to $\rho$ and $\sigma=\rho$, we obtain operators $A^{(j,k)}$ and
$B^{(j,k)}$ intertwining $\rho$ with itself.
It follows from \eqtag{ABrhosigma} that
$$\tilde A^{(j,k)}_{il} = A^{(l,i)}_{kj}, \quad \tilde
B^{(j,k)}_{il} = B^{(l,i)}_{kj}.  \eqno\eq{ABtildeAB}$$
By the Schur lemma, there are complex numbers
$\alpha_{jk}$ and $\beta_{jk}$ such that
$$A^{(j,k)}_{il} = \alpha_{jk} \delta_{il}, \quad B^{(j,k)}_{il}  =
\beta_{jk} \delta_{il},\eqno\eq{ABschur}$$
since $\rho$ is irreducible.
If we sum over $i=l$ in \eqtag{tildeABrhosigma}
we get
$$\tr \, \tilde A^{(j,k)} = \delta_{jk}, \quad \tr\, \tilde B^{(j,k)} =
\delta_{jk}. \eqno\eq{ABtr}$$
Hence, there exists a non-zero intertwining operator $F$ for $\rho$ and
$\rho''$. Since $S$ is invertible,
$\rho$ and $\rho''$ are both irreducible and therefore $F$ is invertible.
So $\rho$ and $\rho''$ are
equivalent corepresentations. Since an intertwining operator
between equivalent irreducible corepresentations
is uniquely determined up to a scalar factor, we conclude from \eqtag{ABtr}
that $\tr(F) \neq 0$ and
$\tr(F^{-1}) \neq 0$. Moreover, there exist complex numbers
$\tilde \alpha_{jk}$ and $\tilde \beta_{jk}$
such that
$$\tilde A^{(j,k)}_{il} = \tilde\alpha_{jk} F_{il}, \quad
\tilde B^{(j,k)}_{il} = \beta_{jk} (F^{-1})_{il}.
\eqno\eq{ABF}$$
Combination with \eqtag{ABtr} yields that
$$\tilde\alpha_{jk} \tr(F) = \delta_{kj}, \quad \tilde\beta_{jk}
\tr(F^{-1}) = \delta_{kj}.
\eqno\eq{Ftr}$$
The  identities \eqtag{schureqone} and \eqtag{schureqtwo} follow
from \eqtag{tildeABrhosigma},
\eqtag{ABF} and \eqtag{Ftr}. This concludes the proof.
\endproof

Let $V$ be a finite-dimensional vector space with inner product
$\innp {\,} {\,} $.
We recall that a linear mapping $T\colon V\to V$ is called
{\defstil positive definite}
if  $T$ is self-adjoint, i.e.\ $T=T^\ast$, and if $\innp Tx x  > 0$ for all
$x\neq 0$. A matrix $(T_{ij})$ is positive definite (with respect to the
canonical
inner product on $\CC^n$) if and only if $T_{ij} = \overline{T_{ji}}$ and
$\sum_{i,j} x_i \bar x_j T_{ji} > 0$ for any $n$-tuple
$(x_1, \ldots, x_n)\neq 0$.
If $T$ is any invertible matrix then ${}^t T \bar T$
is positive definite, where $\bar T = (\overline{T_{ij}})$.
\beginth Proposition Fposdef
{\thstil Let $A$ be a CQG algebra and let $\rho$ be a finite-dimensional
irreducible
unitary corepresentation of $A$ in an inner product space $V$.
Let $F$ be an invertible operator intertwining
$\rho$ and $\rho''$. Then $F$ is a constant multiple
of a positive definite operator
on $V$. It can be uniquely normalized such that $\tr(F) = \tr(F^{-1}) > 0$.
\par}
\endth
\beginproof
Let us fix an orthonormal basis of $V$. Then $\rho$ and $\rho''$
can be viewed as
matrix corepresentations. By \thtag{CQGirredunit} (iii) there is
a unitary matrix
corepresentation $\sigma$ which is equivalent to $\rho'$. So there is an
invertible complex matrix $T$ such that $\sigma T = T \rho'$.
By \eqtag{contragcocoeff}
and the unitarity of $\sigma$ and $\rho$  we have the
identities $$\sigma'_{ij} = \sigma^\ast_{ij} = S(\sigma_{ji}),
\quad \rho'_{ij} = \rho^\ast_{ij}, \quad \rho''_{ij} = S(\rho'_{ji}).$$
It now follows from $\sigma T = T \rho'$ that
$\sigma' \bar T = \bar T \rho$ and
${}^t T \sigma' = \rho'' {}^t T$. Hence
${}^t T \bar T \rho = \rho'' {}^t T \bar T$,
in other words, ${}^t T\bar T$ intertwines $\rho$ and $\rho''$.
Therefore $F$ is a constant multiple of ${}^t T\bar T$. Since
${}^t T\bar T$ is a
positive definite matrix, the first assertion follows.
The second one is trivial.
\endproof
\beginth Theorem haarfaithful
{\thstil Let $A$ be a CQG algebra. Then the Haar functional
$h\colon A\to \CC$ is
positive definite.
\par}
\endth
\beginproof
It follows from \thtag{schurrelnoneq}, \thtag{schurreleq} and
\thtag{Fposdef} that
there exist positive definite matrices $G^\alpha$ such that
$$h((\pi^\alpha_{kl})^\ast \pi^\beta_{ij}) = \delta_{\alpha\beta}
\delta_{lj} G^\alpha_{ik},
\quad \alpha,\beta\in \Sigma.$$
Let $a= \sum_{\alpha,k,l} c^\alpha_{kl} \pi^\alpha_{kl}$
be an arbitrary element of $A$.
Then
$$h(a^\ast a) = \sum_{\alpha,l} \sum_{i,k} \overline{c^\alpha_{kl}}
c^\alpha_{il}
G^\alpha_{ik} \geq 0,$$
since $G^\alpha$ is positive for every $\alpha\in \Sigma$.
Suppose $h(a^\ast a) = 0$. Then $\sum_{ik} \overline{c^\alpha_{kl}}
c^\alpha_{il}
G^\alpha_{ik} = 0$ for all $\alpha$ and $k$. By positive definiteness
of the $G^\alpha$
this implies that all coefficients $c^\alpha_{kl}$ are 0, whence $a=0$.
\endproof
\beginth Remark remarkhaarff
The way we have proved \thtag{haarfaithful} is quite analogous to the
proof of Proposition 3.5 in \Wortan.
\endth
\beginth Proposition  unitarizable
{\thstil Let $A$ be a Hopf $\ast$-algebra on which
there exists a positive definite Haar functional. Then
any finite-dimensional
corepresentation $\pi$ of $A$ is unitarizable and therefore completely
reducible. In particular, the conclusion holds if $A$ is a CQG algebra.
\par}
\endth
\beginproof
Let us denote the corepresentation space of the finite-dimensional
corepresentation $\pi$ by $V$
and  let $\innp {\,} {\,} $ be any inner product
on $V$. We define a new inner product $\innp {\,} {\,} _h$ on $V$ by putting
$$ \innp v w _h = \sum_{(v),(w)} \innp {v_{(1)}} {w_{(1)}}
h(w^\ast_{(2)} v_{(2)}).$$
Indeed, it is clear that $\innp {\,} {\,} _h$ is a hermitian form. Let
$(v_i)$ be an orthonormal basis of $V$ with respect to the inner
product $\innp {\,} {\,} $ and let us write $\pi_{ij}$ for the
matrix coefficients of $\pi$ with respect to this basis. Then
$$\innp v_i v_j _h = \sum_{k,l} \innp v_k v_l  h(\pi^\ast_{lj} \pi_{ki})
= \sum_k h(\pi^\ast_{kj}\pi_{ki}).$$
Hence
$$\innp {\sum_i c_iv_i} {\sum_j c_jv_j} _h  =
\sum_k h((\sum_j c_j\pi_{kj})^\ast
(\sum_i c_i \pi_{ki})) \geq 0, $$
and if the left-hand side equals 0 then $\sum_i c_i \pi_{ki} = 0$ for all $k$
by the positive definiteness of $h$.
Hence $c_k = \varepsilon(\sum_i c_i \pi_{ki})= 0$ for all $k$,
which proves that $\innp {\,} {\,} _h$ is an inner product on $V$.
Using \eqtag{explHaar} one deduces
$$\eqalign{\sum_{(v), (w)} \innp v_{(1)} w_{(1)} _h
w^\ast_{(2)} v_{(2)} &= \sum_{(v),(w)} \innp {v_{(1)}} {w_{(1)}} h(w^\ast_{(2)}
v_{(2)}) w^\ast_{(3)} v_{(3)} \cr
& = \sum_{(v),(w)} \innp {v_{(1)}} {w_{(1)}} h(w^\ast_{(2)} v_{(2)}) \, 1 =
\innp v w _h 1, \cr}$$
in other words, the corepresentation $\pi$ is unitary with respect to the
inner product $\innp {\,} {\,} _h$.
\endproof
\beginth Theorem haareqcqg
{\thstil Let $A$ be a Hopf $\ast$-algebra. Then
there exists a positive definite Haar functional on $A$ if and only if
$A$ is a CQG algebra.
\par}
\endth
\beginproof
The implication $\Leftarrow$ follows from \thtag{haarfaithful}.
Conversely, suppose that there exists a positive definite Haar functional on
the Hopf $\ast$-algebra $A$.
By \thtag{unitarizable} any finite-dimensional
corepresentation of $A$ is unitarizable.
We next claim that every element $a\in A$ occurs as a matrix coefficient of
some finite-dimensional (hence unitarizable) corepresentation of $A$. Indeed,
by the Fundamental Theorem on Coalgebras
(cf.\  \Sweedler,
Th.\  2.2.1, p.\  46) there is a finite-dimensional subcoalgebra
$C$ of $A$ containing $a$. Let us denote the restriction
of $\Delta$ to $C$ by $\pi$. Then clearly $\pi$ is a
corepresentation of $A$ in the finite-dimensional vector space
$C$. To prove our claim, it suffices to exhibit an element $c\in C$
and a linear form
$c'$ on $C$ such that $(c'\ten\id)\circ  \pi(c) = a$. We take
$c= a$ and $c' = \varepsilon_{\vert C}$. It is trivial to check that this
works. This concludes the proof.
\nextsektion
%------------------------------------------------------------------
%\input ndsec4
%------------------------------------------------------------------
\beginsektion 4 {$C^\ast$-algebra completion}
We shall now  show that any CQG algebra can be naturally completed to
a unital $C^\ast$-algebra. Let us recall that a $\ast$-representation
of a $\ast$-algebra $A$ in a Hilbert space $\cal H$ is a $\ast$-algebra
homomorphism of $A$ into the algebra $\cal L(\cal H)$ of bounded operators
on $\cal H$.

Let $A$ be a unital $\ast$-algebra. A seminorm $p$ on $A$ is called a
{\defstil $C^\ast$-seminorm}
if $p(ab) \leq p(a)p(b)$ and  $p(a^\ast a) = p(a)^2$. It then automatically
follows that
$p(a^\ast) = p(a)$. In addition, if $p\neq 0$ then $p(1) = 1$.
A $C^\ast$-seminorm $p$ on $A$ is called a {\defstil $C^\ast$-norm}
if $p(a)=0$ implies
$a=0$. Given a $C^\ast$-norm $p$ on $A$, the completion of  $A$ with respect
to $p$ naturally is a unital $C^\ast$-algebra such that the canonical injection
of $A$ into its completion is a $\ast$-algebra homomorphism.
If $\pi\colon A\to B$ is a $\ast$-algebra homomorphism of $A$ into a
$C^\ast$-algebra $B$,
then the mapping $a\mapsto ||\pi(a)||$
is a $C^\ast$-seminorm on $A$. In particular, every
$\ast$-representation of $A$ gives rise to a $C^\ast$-seminorm.

\beginth Lemma irrepstarnorm
{\thstil Let $A$ be a $\ast$-algebra with a $C^*$-norm $p$.
For any $a\in A$, there exists an
irreducible  $\ast$-representation $\pi\colon A\to {\cal L}({\cal H})$
of $A$ in some Hilbert space $\cal H$ such that $||\pi(a)|| = p(a)$.
\par}
\endth

\beginproof
The result will follow from the corresponding statement for the $C^*$-algebra
completion $\tilde A$ of $A$.
For a proof in that case see for instance \Arv\
(Corollary to Theorem 1.7.2, p.\ 34).
\endproof

\beginth Lemma starext
{\thstil Let $A$ be a Hopf $\ast$-algebra. Let $\pi$ be an algebra homomorphism
of $A$ into the algebra of linear operators on some prehilbert space $V$
such that $(\pi(a)v,w)=(v,\pi(a^*)w)$ for all $a\in A$ and all $v,w\in V$.
Let $\rho=(\rho_{ij})$ be a unitary matrix corepresentation of $A$.
Then $\pi(\rho_{ij})$ is a bounded linear operator on $V$ of norm $\le 1$
for all $i,j$. If $A$ is a CQG algebra then $\pi$ can be uniquely
extended to a $*$-representation of $A$ in the Hilbert space completion of $V$.
\par}
\endth

\beginproof
Since $\rho$ is unitary, we have $\sum_k \rho_{kj}^*\,\rho_{kj}=1$.
Hence, for all $v\in V$,
$$
||v||^2=(v,v)=\sum_k(\pi(\rho_{kj}^*\rho_{kj})v,v)=
\sum_k(\pi(\rho_{kj})v,\pi(\rho_{kj})v)\ge||\pi(\rho_{ij})v||^2.
$$
\endproof

\beginth Lemma unitnorm
{\thstil Let $A$ be a Hopf $\ast$-algebra and let $p$ be
a $C^\ast$-seminorm on $A$. If $\rho=(\rho_{ij})$ is a unitary
matrix corepresentation of $A$ then $p(\rho_{ij}) \leq 1$ for all $i,j$.
\par}
\endth
\beginproof
The subset $N = \{a\in A \mid p(a) = 0\}$ is a two-sided $\ast$-ideal in $A$
and the quotient $A/N$ naturally is a $\ast$-algebra. Let $\phi\colon
A\to A/N$ denote the canonical surjection. Then we can put $\bar p(\phi(a)) =
p(a)$,
and $\bar p$ clearly is a $C^\ast$-norm on $A/N$. Fix $i,j$.
By \thtag{irrepstarnorm} there is a $*$-representation $\bar\pi$ of $A/N$
in some Hilbert space $\cal H$ such that
$||\bar\pi(\phi(\rho_{ij}))||=\bar p(\phi(\rho_{ij}))$.
Hence the $*$-representation $\pi=\bar\pi\circ\phi$ of $A$ in $\FSH$
satisfies $\pi(\rho_{ij})=p(\rho_{ij})$.
Now apply \thtag{starext}.
This concludes the proof of our assertion.
\endth

\beginth Theorem  supnorm
{\thstil Let $A$ be a CQG algebra and let $\goth P$
denote the set of $C^\ast$-seminorms on $A$. The set $\goth P$ is non-empty.
For any $a\in A$, the number
$$\Vert a\Vert_\infty = \sup_{p\in\goth P} p(a)\eqno\eq{sup}$$
is finite. The mapping $a \to \Vert a \Vert_
\infty$ is a $C^\ast$-norm. The norm completion $A^\dagger$ of $A$ with
respect to $\Vert \cdot \Vert_\infty$  naturally is a unital $C^\ast$-algebra.
\par}
\endth

\beginproof
Clearly, the mapping $x\mapsto |\eps(x)|$ of $A$ into $\RR$ is a
$C^\ast$-seminorm.
The fact that \eqtag{sup} is finite follows from
\thtag{unitnorm}. It then is clear that $||\cdot||_\infty$ is a
$C^\ast$-seminorm. To show that $\Vert \cdot \Vert_\infty$
actually is a norm, it suffices to exhibit a $C^\ast$-norm on
$A$. Let $h$ denote the Haar functional on $A$. We define an
inner product on $A$ by putting $\innp a b _h = h(b^\ast a)$. It follows
from \thtag{Haarprop} and \thtag{haarfaithful} that all the inner product
axioms are satisfied.
Left multiplication on $A$ defines an algebra homomorphism of $A$ into the
algebra of linear operators on $A$ such that the properties of \thtag{starext}
are satisfied.
Hence, by \thtag{starext}, this algebra homomorphism
can be extended to a $\ast$-representation $\pi$ of $A$ on the Hilbert
space completion ${\cal H}_h$ of $A$ by \thtag{starext}. Clearly, $\pi$ is
faithful and therefore $a\to \Vert\pi(a)\Vert_h$ is a $C^\ast$-norm.
 Here $\Vert \cdot \Vert_h$ denotes the operator norm on the space
of bounded operators on the Hilbert space
${\cal H}_h$.
\endproof
\beginth Remark remsupnorm
The essence of \thtag{supnorm} can be extracted from \Wortan\ (after
Proposition 3.5). N. Andruskiewitsch communicated to us that a detailed proof
of \thtag{supnorm} is contained in a letter of his to A.\ Guichardet
dated June 1993. This proof is also included in \Guichardet.
\endth
For obvious reasons, the norm $||\cdot ||_\infty$ is called the {\defstil
largest
$C^\ast$-seminorm} on $A$.
We call $A^\dagger$ the   {\defstil universal $C^\ast$-algebra completion}
of $A$. It is uniquely determined (up to a unique isomorphism) by the
following universal property:

\beginth Theorem univcompletion
{\defstil Let $A$ be a CQG algebra and let $\iota\colon A\to A^\dagger$
denote the canonical injection of $A$ into its universal $C^\ast$-algebra
completion $A^\dagger$. If $B$ is a $C^\ast$-algebra and $\pi\colon A\to B$
a $\ast$-algebra homomorphism, then there exists a unique $C^\ast$-algebra
homomorphism $\pi^\dagger\colon A^\dagger \to B$ with the property that
$\pi^\dagger\circ\iota = \pi$.
\par}
\endth
\beginproof
This is a direct consequence of the definition of $A^\dagger$
(cf.\ \thtag{supnorm})
 and the fact that $a\mapsto ||\pi(a)||$ is
a $C^\ast$-seminorm.
\endproof
\beginth Proposition irrepsup
{\thstil  Let $A$ be a CQG algebra. For any $a\in A$,
one has $\Vert a \Vert_\infty = \sup_\pi \Vert \pi(a)\Vert$, where
$\pi$ runs through a complete set of irreducible $\ast$-repres\-entations
of $A$.
 \par}
\endth
\beginproof
This follows by applying \thtag{irrepstarnorm} to the $C^\ast$-algebra
completion
$A^\dagger$ of $A$.
\endproof
\beginth Remark extcomultcounit
Let $A$ be a CQG algebra. Counit and comultiplication on $A$ have unique
extensions to $A^\dagger$.  For the counit this follows from \thtag{supnorm},
since
$\varepsilon\colon A\to \CC$ is a one-dimensional $\ast$-representation of $A$.
For the extension of $\Delta$ to $A^\dagger$ we need a suitable $C^\ast$-norm
on the algebraic tensor product $A^\dagger\ten A^\dagger$. We define the
{\defstil injective cross norm} on $A^\dagger\ten A^\dagger$ by setting:
$$||a||_i = \sup_{\pi_1,\pi_2} ||(\pi_1\ten\pi_2) (a)||,
\quad a\in A^\dagger\ten A^\dagger,
\eqno\eq{injectcross}$$
where $\pi_1$ and $\pi_2$ run through the set of $\ast$-representations
of the $C^\ast$-algebra $A^\dagger$. The mapping $a\mapsto ||a||_i$ clearly
is a $C^\ast$-norm on
$A^\dagger\ten A^\dagger$. Now $a\mapsto (\pi_1\ten \pi_2)(\Delta(a))$ is a
$\ast$-representation of $A$ for any two $\ast$-representations $\pi_1$ and
$\pi_2$ of $A$, so $||(\pi_1\ten\pi_2)(\Delta(a))||\leq ||a||_\infty$,
whence $||\Delta(a)||_i \leq ||a||_\infty$. This implies that $\Delta$
extends to a
continuous mapping of $A^\dagger$ into the completion of
$A^\dagger\ten A^\dagger$
with respect to $||\cdot ||_i$.
\endth

\beginth Remark commutcqg
Let  $A$ be a commutative CQG algebra.
Then the irreducible $\ast$-representations of $A$ are
exactly its one-dimensional $\ast$-representations, in other words,
the points of the compact group $G=G(A)$ corresponding to $A$.
So $||a|| = \sup_{x\in G} ||a(x)||$,
where we view $a\in A$ as a representative function on $G$. By the
Peter-Weyl theorem
$A^\dagger$ is
isometrically isomorphic to the $C^\ast$-algebra of
continuous functions on the group
$G$.
\endth
\nextsektion
%-------------------------------------------
%\input ndsec5
%--------------------------------------------
\beginsektion 5 {Comparison with other literature}
(a) {\sl Woronowicz \Worcomp, \Wortan, \Worrem}
\smallskip\noindent
Woronowicz, in his influential 1987 paper \Worcomp,
gives the following definition of a {\defstil compact matrix quantum group}
(originally called {\defstil compact matrix pseudogroup}).
It is a pair $(B,u)$, where $B$ is a unital $C^\ast$-algebra and
$u=(u_{ij})_{i,j=1,\ldots,N}$ is an $N\times N$ matrix with entries in $B$,
such that the following properties hold.
\smallskip
\item{1)}
The unital $\ast$-subalgebra $A$ of $B$ generated by the entries of $u$
is dense in $B$.
\item{2)}
There exists a (necessarily unique)
$C^\ast$-homomorphism $\Delta\colon B\to B\ten B$ such that
$\Delta(u_{ij})=\sum_{k=1}^N u_{ik}\ten u_{kj}$.
\item{3)}
There exists a (necessarily unique)
linear anti-multiplicative mapping $S\colon A\to A$ such that
$S\circ \ast\circ S\circ \ast=\id$ on $A$ and
$\sum_{k=1}^N S(u_{ik})\,u_{kj}=\delta_{ij}\,1=
\sum_{k=1}^N u_{ik}\,S(u_{kj})$.
\smallskip\noindent
In his note \Worrem\  Woronowicz shows that, instead of 3),
we may equivalently require:
\item{3')}
The matrix $u$ and its transpose are invertible.
\medskip\noindent
Woronowicz now essentially shows (cf.\ \Worcomp\ Prop.\ 1.8)
that there exists a (necessarily unique)
$\ast$-homomorphism $\ep\colon\ A\to\CC$ such that $\eps(u_{ij})=\delta_{ij}$
and that $A$ becomes a Hopf $\ast$-algebra with comultiplication $\Delta$,
counit $\eps$ and antipode $S$. In \Worcomp\  the notation
$A,\FSA,\Phi,e,\kappa$
is used instead of our $B,A,\De,\ep,S$, respectively.
Note that the above $\ast$-algebra $A$ is very close to what we have defined as
a CMQG algebra (cf.\ \thtag{CQGirredunit}). However, it is not postulated
and not yet obvious in the beginning of \Worcomp\ that the corepresentations
$u$ and $u'$ are unitarizable.

A central result in the paper (see \Worcomp\ Theorem 4.2)
is the existence of a {\defstil state}
(normalized positive linear functional) $h$ on the $C^\ast$-algebra $B$
such that $(h\ten\id)\circ \Delta(a)=h(a)\,1=(\id\ten h)\circ \Delta(a)$ for
all $a\in B$.
This state is necessarily unique and it is faithful on $A$.
Then $h$ may be called the Haar functional.

Woronowicz \Worcomp\ \S2
defines a {\defstil representation} of the
compact matrix quantum group $(B,u)$ on a finite-dimensional vector space $V$
as a linear mapping
$t\colon V\to V\ten B$ such that $(t\ten\id)\circ t=(\id\ten\De)\circ t$.
If $t(v)=0$ implies $v=0$ then the representation is called
{\defstil non-degenerate}
and if $t(V)\i V\ten A$ then the representation is called {\defstil smooth}.
A smooth representation is non-degenerate iff $(\id\ten\ep)\circ t=\id$.
Thus corepresentations of $A$ on finite-dimensional vector spaces, as defined
in \S1 of the present paper, correspond to nondegenerate smooth
representations  of $(B,u)$ in \Worcomp.

As a consequence of the existence of the Haar functional, it is shown
in \Worcomp\ Theorem 5.2, Prop.\ 3.2 that nondegenerate smooth
representations of $(B,u)$ are unitarizable. This implies that the
dense $\ast$-algebra $A$ of $B$ is a CMQG algebra.

Conversely, if we start with a CMQG algebra $A$
with fundamental corepresentation
$u$ as in Proposition \thtag{CQGirredunit}
then we have
shown the existence of a positive definite Haar functional $h$ on $A$
(cf.\ Theorem \thtag{haarfaithful}) without using $C^\ast$-algebras,
and we have
next obtained a $C^\ast$-completion $A^\dagger$ of $A$ by making essential
use of the existence of a positive definite Haar
functional (cf.\ section 4).
Then it is clear that the pair $(A^\dagger,u)$
is a compact matrix quantum group in the sense of Woronowicz.
However, the $C^\ast$-algebra $A^\dagger$ possesses the universal property
\thtag{univcompletion}
but this is not necessarily
the case with the compact matrix quantum groups $(B,u)$ of Woronowicz,
since the norm induced
by $B$ on $A$ may not be the largest $C^\ast$-seminorm on $A$.
Accordingly, the counit $\ep\colon \FSA\to\CC$ does not necessarily have
a continuous extension to a linear functional on $A$
(cf.\ \Worcomp\  (second Remark to Prop.\ 1.8)).

In his paper \Wortan, Woronowicz starts with a compact matrix quantum group
$(B,u)$ in the sense of \Worcomp, then constructs out of its
finite-dimensional unitary representations a so-called complete concrete
monoidal $W^\ast$-category (see \Wortan\  Theorem 1.2) and next
constructs from any such category a compact matrix quantum group
$(A^\dagger,u)$.
Then $A^\dagger$ is not necessarily isomorphic to $A$, but it has the universal
property \thtag{univcompletion}
with respect to the CMQG algebra $A$ generated by the entries of $u$.
The relation between CQG algebras and monoidal $W^\ast$-categories is
much closer (cf.\ \msdth).

Both in \Worcomp\ and in the present paper there is a similar key result
\Worcomp\ (last statement of Theorem 5.4 resp.\
Proposition \thtag{Fposdef}). We got the idea of the statement and proof
of Proposition \thtag{Fposdef} from \Worcomp, but in the present paper,
different from \Worcomp,
the positivity and faithfulness of the Haar functional on $\FSA$ is a corollary
rather than a prerequisite.

\newtheme\noindent
(b) {\defstil Woronowicz  \Worprep\ and S. Wang \Wang, \Wangdis}
\smallskip\noindent
Woronowicz \Worprep\ defines a {\defstil compact quantum group}
as a pair $(B,\Delta)$, where $B$ is a separable unital $C^\ast$-algebra and
$\Delta\colon B\to B\ten B$ is a $C^\ast$-homomorphism, such that the following
properties hold.
\smallskip
\item{1)}
$(\Delta\ten\id)\circ\Delta=(\id\ten\Delta)\circ\Delta$.
\item{2)}
$\Span\{(b\ten 1)\,\Delta(c)\mid b,c\in B\}$
and $\Span\{(1\ten b)\,\Delta(c)\mid b,c\in B\}$
are dense subspaces of $B\ten B$.
\smallskip\noindent
In particular, if $(B,u)$ is a compact matrix pseudogroup as defined in
\Worcomp\  and if $\Delta$ is the corresponding comultiplication
then $(B,\Delta)$ is a compact quantum group as just defined.
Conversely, it is shown in \Worprep\  that, if $(B,\Delta)$ is a compact
quantum group and if $A$ is the set of all linear combinations
of matrix elements of finite-dimensional unitary representations of
$(B,\Delta)$
then $A$ is a dense $\ast$-subalgebra
of $B$ and $A$ is a Hopf $\ast$-algebra.
The existence of a Haar functional is also shown.
It is observed that the representation theory as developed in \Worcomp\
can be formulated in a similar way for compact quantum groups.

It is pointed out in Wang (\Wang\ Remark 2.2) that the results of
\Worprep\ remain true if separability of the $C^\ast$-algebra $B$ is no longer
required, but if it is assumed instead that the $C^\ast$-algebra $B$
has a faithful state.
This observation would imply that a compact quantum group $(B,\Delta)$ in the
sense of Wang gives rise to a CQG algebra $A$ ($A$ being
constructed from $B$ as in the previous paragraph), and that conversely each
CQG algebra $A$ would give rise to a compact quantum group $(B,\Delta)$
($B$ being completion of $A$ with respect to maximal $C^\ast$-seminorm),
provided $B$ has a faithful state.

In \Wangdis, Wang defines the notion of (non-commutative) Krein algebra, which
is
essentially equivalent to our notion of CQG algebra.
\newtheme\noindent
(c) {\defstil Effros \& Ruan \Effros}
\smallskip\noindent
In different terminology,
CQG algebras were earlier introduced by Effros \& Ruan \Effros.
They defined these algebras as cosemisimple Hopf algebras with
a so-called standard $\ast$-operation and
they called these structures {\defstil discrete quantum groups}.
This name was motivated by the fact that special examples of these algebras
are provided by the group algebra of a discrete group,
while the name CQG algebra comes from the class of examples where we
deal with the algebra
of representative functions on a compact group.
In the final section of \Effros\  the authors define a
{\defstil compact quantum group} as a natural generalization of the compact
matrix quantum groups defined in \Worcomp. Their definition involves
a unital $C^\ast$-algebra $B$ with a dense unital $\ast$-subalgebra $A$,
where $A$ is a CQG algebra (in the terminology of the present paper)
and the comultiplication on $A$ extends continuously to $B$.
Conversely, they show that a CQG algebra $A$ gives rise to a compact
quantum group according to their definition.
This involves a $C^\ast$-completion, for which a Haar functional $h$ on $A$
is needed. This Haar functional is obtained in a way very different from
the method in the present paper. The authors first show the existence
of a left Haar functional $\phi$ on a certain subspace
of the linear dual of $A$. Then $h$ is constructed in terms of $\phi$.
For a detailed comparison of \Effros\ with the results in the present paper,
see \Kustermans\ (section 6).
\nextsektion
%---------------------------------------------------
%\input ndbiblio
%------------------------------------------------
\def\bibitem#1#2\par{\vskip0.05cm\item{#1} #2.\par}
\font\refstil=cmsl10
\beginsektion 6 {References}
{\parindent=1.2cm
\bibitem \Arv Arveson, W., {\refstil An Invitation to $C^\ast$-algebra,}
Graduate Texts in Mathematics 39,
   Springer-Verlag, 1976\par
\bibitem \Bonneau Bonneau, P., M.\ Flato, M.\ Gerstenhaber, G.\ Pinczon,
{\refstil The hidden group structure
of quantum groups: strong duality, rigidity and preferred deformations,}
Comm.\ Math.\ Phys.\ 161 (1994), 125-156\par
\bibitem \BourbakiAlgeight Bourbaki, N., {\refstil El\'ements de
Math\'ematique,} Livre II,
       {\refstil Alg\`ebre,} Ch.\ 8, Hermann, 1958\par
\bibitem \msdth Dijkhuizen, M.S., {\refstil On compact quantum groups and
quantum homogeneous spaces,}
Dissertation, University of Amsterdam, 1994\par
\bibitem \Effros Effros, E.G., Z.\ Ruan, {\refstil Discrete quantum groups I.
The Haar measure,} preprint
(University of California at Berkeley), 1993, to appear in Int.\ J.\ Math.\par
\bibitem \Gerst Gerstenhaber, M., A.\ Giaquinto, S.D.\ Schack, {\refstil
Quantum symmetry,} in:
{\refstil Quantum Groups,}, ed.\ P.P.\ Kulish, Lecture Notes in Mathematics
1510, Springer-Verlag, 1992,
9--46\par
\bibitem \Guichardet Guichardet, A., {\refstil Introduction aux groupes
quantiques,} preprint, Ecole
Polytechnique, Palai\-seau, August 1993\par
\bibitem \Koorlect Koornwinder, T.H., {\refstil General compact quantum
groups, a tutorial,} Report 94-06,
Math. Preprint Series, Dept.\ of Math.\ and Comp.\ Sci., Univ. of Amsterdam,
1994, to appear as part of
{\refstil Compact quantum groups and $q$-special functions,} in: {\refstil
Representations of Lie Groups and
Quantum Groups,} ed.\ M.\ Picardello, Longman\par
\bibitem \Kustermans Kustermans, J., {\refstil Two notions of discrete quantum
groups,} preprint K.U.\ Leuven,
1994\par
\bibitem \Maclane MacLane, S.,  {\refstil Categories for the Working
Mathematician,} Graduate Texts in
Mathematics 5, Springer-Verlag, 1971\par
\bibitem \FRT Reshetikhin, N.Yu., L.A.\  Takhtadjan, L.D.\ Faddeev,
{\refstil Quantization of Lie groups and Lie
algebras,} Leningrad Math.\ J.\ 1 (1990), 193--225\par
\bibitem \Sweedler Sweedler, M.E., {\refstil Hopf Algebras}, Benjamin, 1969\par
\bibitem \Vaksoib Vaksman, L.L., Ya.\ S.\ Soibelman, {\refstil Algebra of
functions on the quantum group
$SU(2)$,} Funct.\ Anal.\ Appl.\ 22 (1988), no.\ 3, 170--181\par
\bibitem \Wang Wang, S., {\refstil General constructions of compact quantum
groups,} preprint (University of California
at Berkeley), 1993\par
\bibitem \Wangdis Wang, S., {\refstil General constructions of compact quantum
groups,} dissertation, Berkeley, 1993\par
\bibitem \Worcomp Woronowicz, S.L., {\refstil Compact matrix pseudogroups},
Comm.\ Math.\ Phys.\
          111 (1987), 613--665\par
\bibitem \Wortan Woronowicz, S.L., {\refstil Tannaka-Krein duality for compact
matrix pseudogroups. Twisted
            $SU(N)$ groups,} Invent.\ Math.\ 93 (1988), 35--76\par
\bibitem \Worrem Woronowicz, S.L., {\refstil A remark on compact matrix
quantum groups,}
       Lett.\ Math.\ Phys.\ 21 (1991), 35--39\par
\bibitem \Worprep  Woronowicz, S.L., {\refstil Compact quantum groups,}
preprint (University of Warsaw),
version August 1, 1993\par
\par}
\nextsektion

\bye